# Combining configuration and recommendation to define an interactive product line configuration approach

CAMILLE SALINESI, RAOUIA TRIKI, RAUL MAZO
CRI
Panthéon Sorbonne University
90, rue de Tolbiac, 75013 Paris
FRANCE
{camille.salinesi, raouia.triki}@univ-paris1.fr, raulmazo@gmail.com

*Abstract:* This paper is interested in e-commerce for complex configurable products/systems. In e-commerce, satisfying the customer needs is a vital concern. One particular way to achieve this is to offer customers a panel of options among which they can select their preferred ones. While solution exists, they are not adapted for highly complex configurable systems such as product lines. This paper proposes an approach that combines two complementary forms of guidance: configuration and recommendation, to help customers define their own products out of a product line specification. The proposed approach, called interactive configuration supports the combination by organizing the configuration process in a series of partial configurations where decisions are made by the recommendation. This paper illustrates this process by applying it to an example with the content based method for recommendation and the a priori configuration approach.

*Key-Words:* configuration, recommendation, product line, feature

## 1 Introduction

In e-commerce, one particular way to let customers define their needs is to offer them a panel of options among which they can select their preferred ones. The systems engineering community has shown that it is then possible to produce customized products (such as software or systems) based on methods, techniques and tools engineering. A Product Line (PL) is a family of products[*] sharing common characteristics and satisfying the needs of a particular mission [29]. In a Product Line, products are not individually explicitly predefined in advance. On the contrary they are produced consistently with the customers' requirements based on the specification of the PL. In order to achieve this, marketing and engineers must define upstream models that specify reusable artifacts and the constraints to prescribe correct combinations. These models are interpreted by configurators to decide whether a requested configuration, specified as a combination of reusable artifacts or through more complex requirements [8], is correct or not.

The first issue of Web configurators is performance [37]. Indeed, as soon as a customer makes a decision (e.g. to require or reject reusable artifacts), he/she wants to find out what the consequences of the decisions are. From a marketing perspective, it is unpleasant for the customer to wait for several seconds to know whether his/her requirements are correct or not in terms of configuration. This issue is extremely difficult to solve at the industrial level, because the computational complexity of computing PL configuration grows quadratically [37]. The second issue is guidance. In practice, choosing from a wide range of options is quickly difficult for the customer who doesn't know where to start, or which alternative choose. Besides each times a customer makes a decision, this can be contradictory with previous decisions, or have a negative impact on downstream decisions. The customer will be disappointed to see that his choices are not correct, with the risk that he/she ultimately turns to a competitor. Therefore, it is crucial to guide the customer in the PL configuration process.

---
[*]Products: Products can be software, socio technical systems, or complex products such as car, train, plane, etc.

This raises the question of how to increase scalability in the guidance of the configuration process, while maintaining a reasonable performance?

E-commerce systems (such as Amazon) already offer guidance to choose in a very large collection of products using recommendation techniques. However, these recommendation techniques are not adapted to complex systems such as product lines, configurable software, or composite systems with a large number of options. Indeed, although recommendation techniques can deal with many options to decide upon, they do not take into account the constraints between them. We are thus faced to a situation where configuration systems and recommendation systems solve two parts of our problem, but not in an integrated way.

Our research goal is to explore the combination of two complementary forms of guidance: recommendation and configuration. We call the approach that we have developed interactive configuration. The aim of interactive configuration is to inform the customer in real time about desirable/possible/unattainable features according to his/her choices, as well as to suggest decision to focus on, and what choice to make by reasoning with known configurations.

We believe that four key issues should be handled to solve the research question:
- What shall the recommendation apply to? (options, alternatives, requirements, etc)
- Which recommendation technique shall be used? (collaborative filtering, content based filtering, etc)
- What type of data should be used for recommendation? (experiences of past configurations, profiles, contextual data, etc)
- How to combine configuration and recommendation?

In this paper, we propose to apply the content based method [35] for recommendation to the a priori configuration approach. The recommendation is applied to the product line characteristics and based on the textual data.

The remainder of the paper is organized as follows. First, we introduce a motivating example in Section 2. Then, we describe our approach in Section 3. In Section 4, we present and discuss related work. Finally, Section 5 concludes the paper.

## 2 Background and Motivating Example

### 2.1 Product Line Engineering

A Software Product Line (SPL) is a set of software-intensive systems that share a common and managed set of software reusable artifacts satisfying the specific needs of a particular market segment or mission and that are developed from a common set of core assets in a prescribed way [29].

Pohl et al. define the Software Product Line Engineering (SPLE) as a paradigm to develop software applications using platforms and mass customization [24]. The SPLE paradigm separates two processes:
- Domain engineering: This process is responsible for establishing the reusable platform and thus for defining the commonality and the variability of the product line. The platform consists of all types of software artefacts (requirements, design, realisation, tests, etc.). It is composed of five key sub-processes: product management, domain requirements engineering, domain design, domain realisation, and domain testing [24].
- Application engineering is the process of software product line engineering in which the applications of the product line are built by reusing domain artefacts and exploiting the product line variability. It is composed of the following sub-processes: application requirements engineering, application design, application realisation, and application testing [24].

Product line models are commonly used to define the valid combinations of reusable artifacts in a product line. Not all artifacts are compatible. The main purposes of product line models are (i) to capture commonalities and variabilities; (ii) to represent dependencies between the reusable artifacts; and (iii) to determine combinations of artifacts that are allowed and disallowed in the PLM. The artifacts can be viewed as reusable requirements, components, functions, etc. Due to the number of system views (requirements, architectural components, processes, etc.) that can be represented by means of PLMs, many modeling formalisms have been proposed in literature. In this paper we use the Feature Models notation, one of the most popular languages used to represent PLMs.

## 2.2 Introduction to Feature Models

FMs were first introduced in 1990 as a part of the Feature-Oriented Domain Analysis (FODA) method [11], as a way to represent the commonalities and variabilities of SPLs. In this method, a *feature* is a prominent or distinctive user-visible aspect, requirement, quality, or characteristic of a software system [11]. A *Feature Model (FM)* defines the valid combinations of features in a software product line. Several extensions have been proposed to improve and enrich their expressiveness. Two of these extensions are cardinalities [25], [7] and attributes [31], [36]. Although there is no consensus on a notation to define attributes, most proposals agree that an attribute is a variable with a name, a domain and a value. Note that the value of attributes is not specified in the product line model. Instead, the value of each attribute is assigned for each particular configuration, (when these attributes are attached to features that belong to the configurations). In this paper, we are interested into these two extensions, illustrated in the Figure 1.

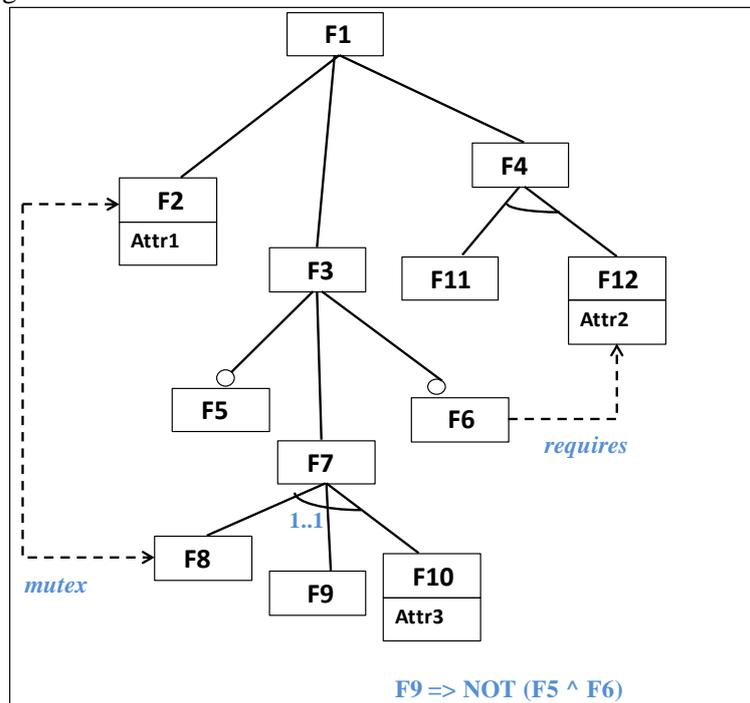

Fig.1 Feature Model

The semantic of the FM's relationships in Figure 1 can be summarized as follows:

*Mandatory:* Given two features F1 and F2, where F1 is the father of F2, a mandatory relationship between F1 and F2 means that if F1 is selected in a product, then F2 must be selected too, and vice versa.

*Optional:* Given two features F1 and F2, where F1 is the father of F2, an optional relationship between F1 and F2 means that if F1 is selected in a product, then F2 may be selected or not. However, if F2 is selected then F1 must also be selected. For example, in the Fig.1, F5 and F6 are optional.

*Requires:* Given two features F1 and F2, a relationship F1 *requires* F2 means that if F1 is selected in a product then F2 has to be selected as well. Additionally, it means that F2 can be selected even when F1 is not selected. For example, in the Fig.1, F6 requires F12.

*Exclusion:* Given two features F1 and F2, a relationship F1 *excludes* F2 means that F1 and F2 cannot be selected in the same product. For example, in the Fig.1, F2 excludes F8 and F8 excludes F2: the *mutex* relation.

*Group cardinality:* A group cardinality is an interval denoted <n..m>, with n as lower bound and m as upper bound limiting within a group of features the number of features that can be part of a product. All the features in the group must have the same parent feature, and none can be selected if the parent is not itself selected. For example, in the Fig.1, <1..1> is a group cardinality.

## 2.3 Running Example

As a running example, we illustrate FMs and our work with the example of the DELL Laptop/Notebook Computers [30]. In this example, represented in Fig. 2, every DELL laptop/notebook has seven mandatory features which are one *Product category*, one *Operating system*, one *Hard drive*, one *Optical drive*, one *Laptop weight*, one *Memory*, one *Processor* and one *Price*. Each of these mandatory feature has child features that are related in a [1..1] group cardinality. For instance, in Fig. 2, *UltraLight*, *Light* and *DesktopReplacement* are related in [1..1] group cardinality, which means that only one of these options can be selected in a DELL laptop/notebook configuration.

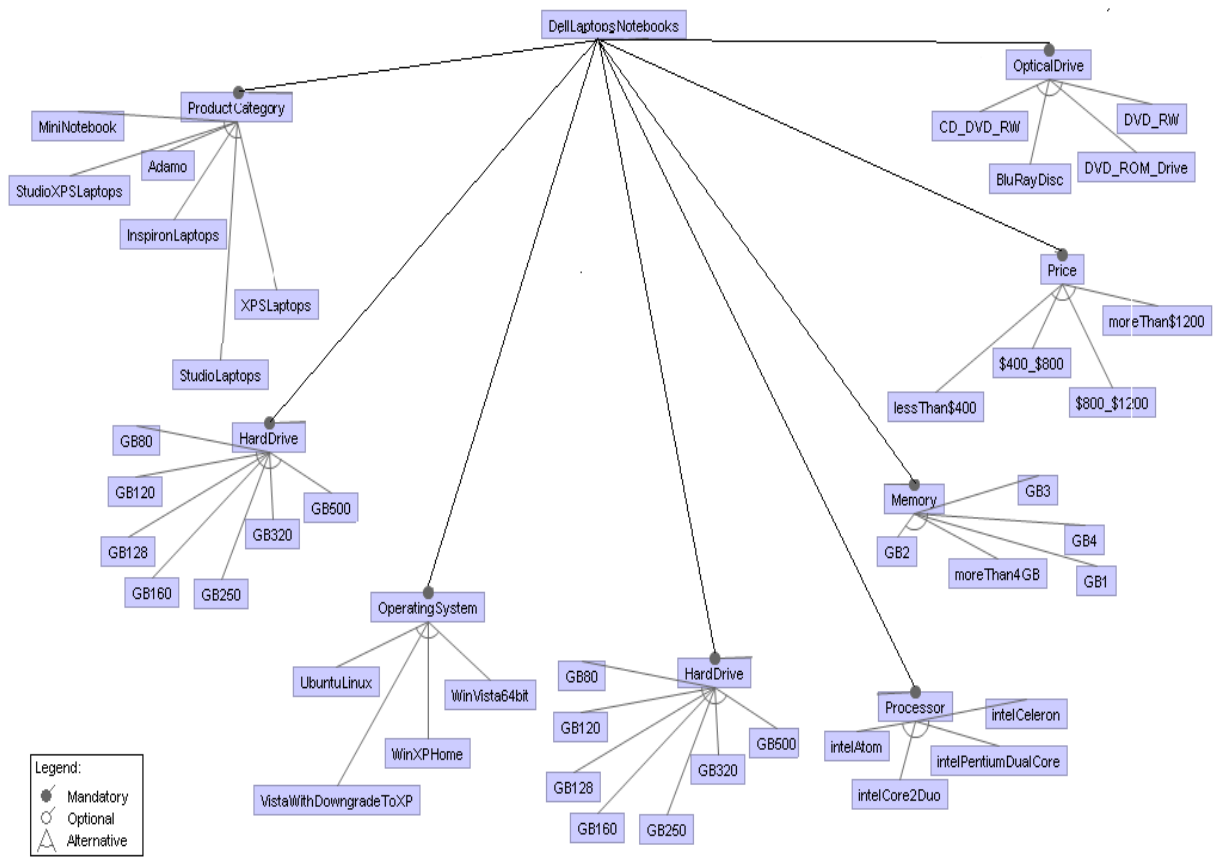

Fig. 2 Feature representation of a DELL Laptops/Notebooks

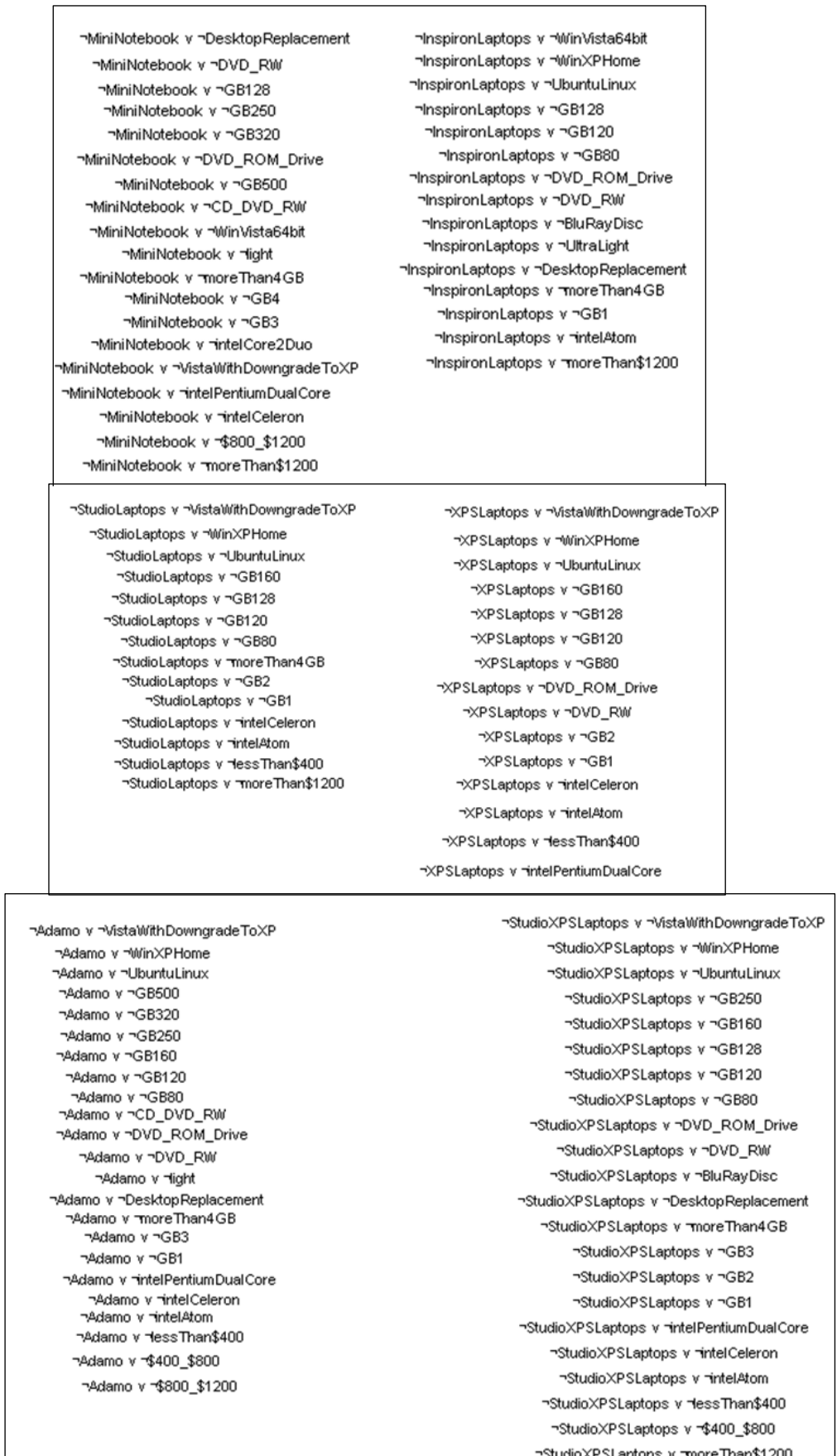

Fig. 3 Constraints of the feature representation of the DELL Laptops/Notebooks

In the example above, features can be classified into subsets. There are features that belong to a subset of functional features and others belong to a subset of nonfunctional features. For instance, Operating *system*, *Hard Drive*, *Optical Drive*, *Memory* and *Processor* are functional features, and *Product Category, Laptop Weight* and *Price* are nonfunctional features.

## 3 Proposed Solution

In an a priori configuration approach, the user selects a configuration by selecting the features, one by one, that he wants to have in the final product. The configurator tests the user's configuration and returns a result that shows if the configuration is correct or not. The test of configuration is done according to the PLM, a feature model in our case. The issue that we want to take in this paper is when the configuration made by the user is not correct.

Our proposal to deal with the aforementioned issue consists in guiding the customer in the configuration by intertwining the recommendation and configuration activities in an iterative way. At each iteration (i) a series of decisions is offered to the customer (ii) the customer makes choices (iii) testing the user configuration (iv) recommendation (v) configuration and constraint propagation (vi) final decision.

The partial configuration strategy was chosen to avoid the issues raised by the a priori configuration strategy. Indeed, in the partial configuration strategy, the customer does not have to decide on all features. Only a subset of features are considered first, then decisions are automatically propagated through the constraints onto the other features.

The approach for recommendation is content based filtering. It is therefore used to help the customer decide on partial configurations. This method treats the recommendations problem as a search for related items [1]. Given the user's purchased and rated items, the algorithm constructs a search query to find other popular items with the similar keywords or subjects [14].

The content of the profile depends on the method used in the analysis of document content. To achieve this, several learning techniques were used, such as analytical techniques taken from textual information retrieval for the recommendation of textual documents [1], [22], [23]. The profile often takes the form of a vector of keywords with weights (1). The weight associated with each word reflects the importance of this term to the user. These words are often extracted using the TF-IDF measure [27] (4). This vector is then compared to that of the document (2). To achieve this, several measurements can be used such as the measurement vector cousin (3).

(1) Profile_Content (c) = $\{(t^c_j, w^c_j)\}$, j=1…k

(2) Content (s) = $\{(t^s_i, w^s_i)\}$, i=1…n

(3) $U(c,s) = \cos(\vec{w_c}, \vec{w_s}) = \sum_{i=1,k} \dfrac{w_{ic}}{\sqrt{\sum_{i=1,k} w^2_{ic}}} \dfrac{w_{is}}{\sqrt{\sum_{i=1,k} w^2_{is}}}$

(4) $W(i, s) = Tf * \log(N/DF_i)$

With:

- $W_{i,c}$ : the weight of the term i in the user profile vector.

- $W_{i,s}$: the weight of the term i in the document vector s.

- Tf: the number of occurrences of the term i in the document s.

- N: the number of document in the collection.

- $DF_i$: the number of documents that contain the term i at least once.

These equations can be implemented into the following algorithm:

```
k=number of terms;
ucs=0 ;
For i= 1 to k do
a1=0;
a2=0;
   For i2=1 to k do
        a1=a1+square(termWeight(term[i2], nameFile1));
        a2=a2+square(termWeight(term[i2], nameFile 2));
End for
W i, c = termWeight (term[i], nameFile 1);
s1=squareRoot(a1);
W i, s = termWeight (term[i], nameFile r2);
s2= squareRoot (a2);
Ucs=ucs+( (wic/s1) * (wis/s2) );
End for
Result=ucs;
```

Our solution recommends a correct configuration to the user that corresponds to his initial configuration if the latter is not correct. Indeed, the content based method uses the definition of existing product which is a combination of features that have already shown correct, then, compares the content of this configuration with the content of the PLM.

The diagram shown in Fig.4 presents an overview of our solution. As the diagram shows it, the process starts with a partial configuration made by the user, and proceeds with larger partial configuration until the configuration is complete.

At each cycle, the content analysis is applied to partial configurations. If the partial configuration is incorrect, this is indicated to the user, who has the choice between improving it, or proceeds following that the recommender will propose a different solution.

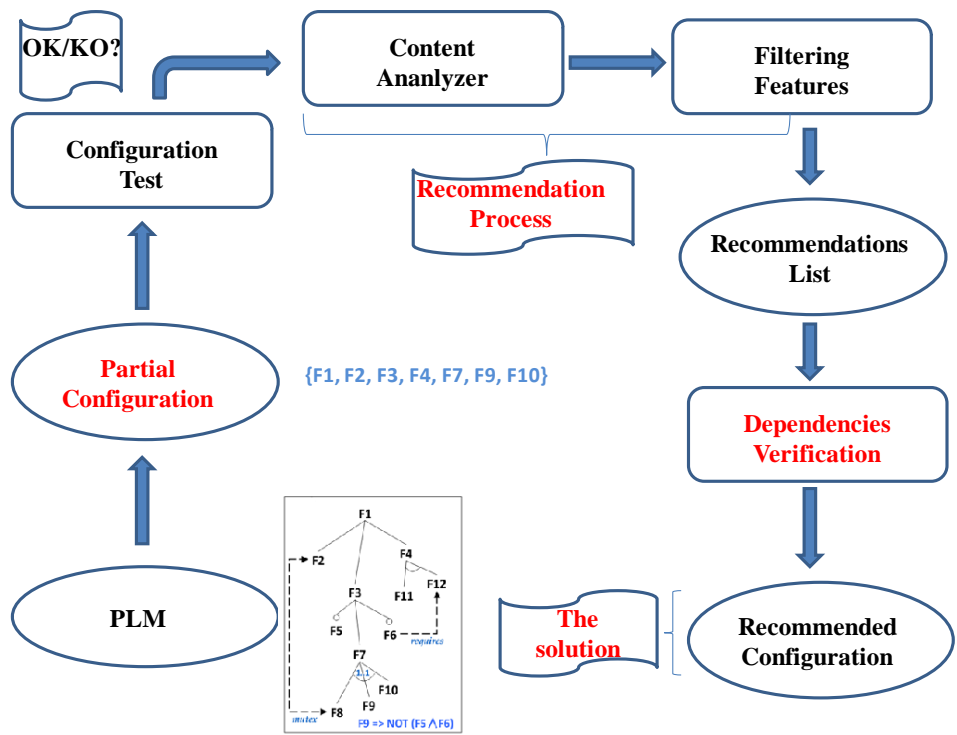

Fig.4 Steps of the solution

Our solution contains the following five steps:
1. Define a partial configuration: The user define a configuration by selecting the product features that satisfy his needs.
2. Testing the user configuration: The configurator tests the user configuration according to the PLM and returns if it is correct or not. In this diagram, we present the PLM as a FODA model.
3. Identify recommendations: we recommend other configurations with features similar to those that the user has chosen. The recommendation process is based on the content based method explained above. The recommendation is applied to a subset of features of the user partial configuration. Indeed, a PLM can be composed of facets making it possible to divide the features into several subsets. For instance, in a PLM, we can distinguish logical features subsets and physical features subsets.
4. Constraint propagation: Verify all the dependencies between the recommended features and the remaining features of the PLM. For example verify the "requires" dependency or the "mutex" dependency between recommended features and the remain features of the PLM.
5. Provide the solution: We recommend a configuration that satisfy the most the user need and that satisfy dependencies of the PLM.

The following example illustrates the application of our solution to the DELL Laptops/Notebooks running example.

1. The configuration selected by the user called *C1* is the following:
- Product Category: Mininotebook
- Operating System: UbuntuLinux
- Hard Drive: 320GB
- Optical Drive: CD_DVD+RW
- Laptop Weight: UltraLight
- Memory: 2GB
- Processor: IntelAtom
- Price: $400_$800

2. This configuration C1 is not correct because:

$$\begin{cases} \text{Not (Mininotebook) v Not (320GB)} \\ \text{And} \\ \text{Not (Mininotebook) v Not (CD\_DVD+RW)} \end{cases}$$

3. Recommendation of other configurations similar to *C1*. The Recommendation is based on **functional features** that are: *Operating System*, *Hard Drive*, *Optical Drive*, *Memory* and *Processor*. There are several possible configurations similar to *C1* (about 2240 configurations). We give only four similar configurations to C1 in order to make a simulation. We called these configurations *C1.1*, *C1.2*, *C1.3* and *C1.4*.

| *C1.1:* | *C1.2:* |
|---|---|
| • Product Category: Mininotebook <br> • Operating System: VistaWithDowngradeToXP <br> • Hard Drive: 160GB <br> • Optical Drive: DVD_ROM_DRIVE <br> • Laptop Weight: UltraLight <br> • Memory: 2GB <br> • Processor: IntelAtom | • Product Category: Mininotebook <br> • Operating System: UbuntuLinux <br> • Hard Drive: 120GB <br> • Optical Drive: BluRayDisc <br> • Laptop Weight: UltraLight <br> • Memory: 1GB <br> • Processor: IntelCore2Duo <br> • Price: $400_$800 |
| • Price: $400_$800 | |
| *C1.3*:<br>• Product Category: Mininotebook<br>• Operating System: UbuntuLinux<br>• Hard Drive: 160GB<br>• Optical Drive: BluRayDisc<br>• Laptop Weight: UltraLight<br>• Memory: 2GB<br>• Processor: IntelAtom<br>• Price: $400_$800 | *C1.4*:<br>• Product Category: Mininotebook<br>• Operating System: WinXPHome<br>• Hard Drive: 120GB<br>• Optical Drive: BluRayDisc<br>• Laptop Weight: UltraLight<br>• Memory: 1GB<br>• Processor: IntelAtom<br>• Price: $400_$800 |

4. Constraint propagation:

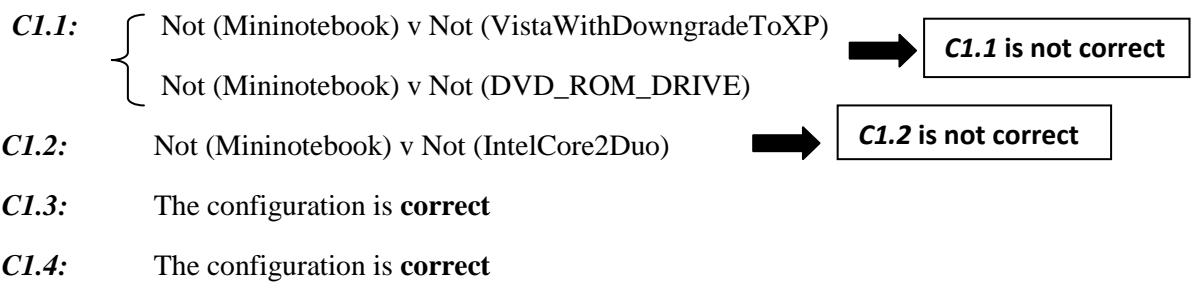

*C1.1*: Not (Mininotebook) v Not (VistaWithDowngradeToXP) → **C1.1 is not correct**
Not (Mininotebook) v Not (DVD_ROM_DRIVE)

*C1.2*: Not (Mininotebook) v Not (IntelCore2Duo) → **C1.2 is not correct**

*C1.3*: The configuration is **correct**

*C1.4*: The configuration is **correct**

5. Provide the solution: Return the two correct configurations *C1.3* and *C1.4*.

# 4 Related Works

## 4.1 Interactive configuration methods

An interactive product configurator is a tool that allows specifying a product according to his specific requirements and the constraints of the PLM to combine these needs. This process can be done interactively, i.e. in a step-wise fashion, and guided, i.e. proposing the resolution of certain requirements before others and automatically proposing a valid solution when there is only one possible choice in the solution space.

To be useful in the e-commerce context, a configurator must be complete (i.e., ensure that no solutions are lost), it must allow order-independent selection/retraction of decisions, give short response times, and offer recommendation to maximize the possibilities to have one satisfactory configuration.

Solution techniques applied to the configuration problem have been compared by Benavides et al. [5], Hadzic & Andersen [9], and Hadzic et al.[10]. They mainly distinguish approaches based on propositional logic on the one hand and on constraint programming on the other hand.

When using propositional logic based approaches, configuration problems are restricted to logic connectives and equality constraints [10], [32]. Arithmetic expressions are excluded because of the underlying solution methods.

These approaches perform in two steps. First, the feature model is translated into a propositional formula. In the second step the formula is solved by appropriate solvers, in particular SAT solvers, as disused by Janota [12], and BDD-based solvers [10], [33]. BDD-based solvers translate the propositional formula into a compact representation, the BDD (Binary Decision Diagram). While many operations on BDDs can be implemented efficiently, the structure of the BDD is crucial as a bad variable ordering may result in exponential size and, thus, in memory blow up. Even if several heuristics are used to improve the use of BDD solvers in the context of PLs [19], these heuristics are not exploitable in industry, they are not even used in the author's web site [18] for heuristic's instability reasons.

Feature models can be naturally mapped into constraint systems in order to reason (e.g., configuration) on them, in particular into CSPs as presented by Benavides et al. [3], [4], [5], Subbarayan et al. [32], and Van Deursen & Klint [34]; into Constraint Programs over finite domains CPs as presented by Mazo et al. [15], [16] and Salinesi et al. [26], and into Constraint Logic Programs (CLPs) as presented by Mazo et al. [17].

Benavides et al. [5] compare the approaches sketched above, particularly with respect to performance and expressiveness or supported operations. They point out that CSP-based approaches, in contrast to others, can allow extra functional features [11] and, in addition, arithmetic and optimization. Furthermore, they state that "the available results suggest" that constraint-based and propositional logic-based approaches "provide similar performance", except for the BDD-approach which "seems to be an exception as it provides much faster execution times", but with the major drawback of BDDs having worst-case exponential size.

Extended feature models with numerical attributes, arithmetic, and optimization are denominated as an important challenge in interactive configuration by Benavides et al. [5]. Our approach tackles this challenge. The main idea is to follow the constraint-based approach, while using recommendation techniques and selection order heuristics in order to satisfy the characteristics of a good configurator in the e-commerce context.

### 4.2 Recommendation

The recommendation algorithms are mainly based on filtering techniques. The idea is to propose customers the products according to predictions made with respect to their preferences. If the choice is made from products that the customer has chosen, it is called "content-based" filtering [13], [21], where made from products that were purchased from other customers, it is called "collaborative" filtering [14], [20].

For a collaborative filtering system, there are two types of algorithms for calculating prediction: memory based filtering algorithms and model based filtering algorithms [6]. Model-based collaborative filtering algorithms provide item recommendation by first developing a model of user ratings. Algorithms in this category take a probabilistic approach and envision the collaborative filtering process as computing the expected value of a user prediction, given his ratings on other items [28]. The model building process is performed by different machine learning algorithms such as Bayesian network and clustering approaches. The Bayesian network model [6] formulates a probabilistic model for collaborative filtering problem. The clustering model treats collaborative filtering as a classification problem [2], [6]. Memory-based algorithms utilize the entire user-item database to generate a prediction. These systems employ statistical techniques to find a set of users, known as neighbors, that have a history of agreeing with the target user. Once a neighborhood of users is formed, these systems use different algorithms to combine the preferences of neighbors to produce a prediction for the active user [28].

## 5 Conclusion

In this paper, we have presented the interactive configuration approach. It is a combined approach of configuration and recommendation of product lines. We have applied the recommendation onto the a priori configuration approach. The recommendation algorithm used in our approach is the content based method. The recommendation process focuses on a features subset of the user configuration and not on all the features. This process allows recommending to the user other configurations similar to that he/she specified at the beginning, focusing on a subset of features that we judge the most important.

## References

[1] Balabanovic M., Shoham Y. Combining Content-based and collaborative recommendation. Communications of the ACM, 40 (3), March 1997.

[2] Basu C., Hirsh H., Cohen W. Recommendation as classification: Using social and content-based information in recommendation. In Proceedings of the Fifteenth National Conference on Artificial Intelligence, pp. 714-720.
[3] Benavides D., Ruiz-Cortés A., and Trinidad P. Using constraint programming to reason on feature models. In The Seventeenth International Conference on Software Engineering and Knowledge Engineering, SEKE 2005, pages 677–682, 2005.
[4] Benavides D., Segura S., Trinidad P., and Ruiz-Cortés A. FAMA: Tooling a framework for the automated analysis of feature models. In Proceeding of the First International Workshop on Variability Modelling of Software-intensive Systems (VAMOS), pages 129–134, 2007.
[5] Benavides D., Segura S., Ruiz-Cortés A. Automated analysis of feature models 20 years later: A literature review. Information Systems, 35(6):615-636, 2010.
[6] Breese J. S., Heckerman D. and Kadie C. Empirical Analysis of Predictive Algorithms for Collaborative Filtering, Uncertainty in Artificial Intelligence, 1998.
[7] Czarnecki K., Helsen S., and Eisenecker U.W. Formalizing cardinality-based feature models and their specialization. Software Process: Improvement and Practice, 10(1):7–29, 2005.
[8] Djebbi O. L'ingénierie des exigencies par et pour les lignes de produits. Université Paris1 Panthéon-Sorbonne, France, PhD Thesis, 2011.
[9] Hadzic T., Andersen H. R. An introduction to solving interactive configuration problems. Technical Report TR-2004-49, The IT University of Copenhagen, 2004.
[10] Hadzic T., Subbarayan S., Jensen R. M., Andersen H. R., Moller J., Hulgaard H. Fast backtrack-free product configuration using a precompiled solution space representation. In International Conference on Economic, Technical and Organizational aspects of Product Configuration Systems (PETO), pages 131-138, 2004.
[11] Kang K., Cohen S., Hess J., Novak W., and Peterson S. Feature–Oriented Domain Analysis (FODA) Feasibility Study. Technical Report CMU/SEI-90-TR-21, Software Engineering Institute, Carnegie Mellon University, November 1990.
[12] Janota M. Do SAT solvers make good configurators? In First Workshop on Analyses of Software Product Lines (ASPL), September 2008.
[13] Lang K. NewsWeeder: learning to filter netnews. In Proceedings of the 12[th] International Conference on Machine Learning, pages 331-339. Morgan Kaufmann publishers Inc.: San Mateo, CA, USA, 1995.
[14] Linden G., Smith B., York J. Amazon.com recommendations: item-to-item collaborative filtering. IEEE Internet Computing, 7(1):76-80, 2003.
[15] Mazo R., Grünbacher P., Heider W., Rabiser R., Salinesi C., Diaz D. Using constraint programming to verify DOPLER variability models. VaMoS 2011: 97-103.
[16] Mazo R., Salinesi C., Diaz D., Lora-Michiels A. Transforming Attribute and Clone-enabled Feature Models into Constraint Programs over Finite Domains. ENASE 2011: 188-199.
[17] Mazo R., Lopez-Herrejon R. E., Salinesi C., Diaz D., Egyed A. Conformance Checking with Constraint Logic Programming: The Case of Feature Models. COMPSAC 2011: 456-465
[18] Mendonça, M., Branco, M., Cowan, D. S.P.L.O.T.: software product lines online tools. In OOPSLA Companion. ACM, (2009) http://www.splot-research.org
[19] Mendonça, M., Wasowski, A., Czarnecki, K. SAT–based analysis of feature models is easy. In Proceedings of the Sofware Product Line Conference (2009).
[20] Miller B. N., Albert I., Lam S. K., Konstan J. A., Riedl J. Movielens unplugged : Experiences with an occasionally connected recommender system. In Proceedings of ACM 2003 Conference on Intelligent User Interfaces (IUI'03). ACM, 2003.
[21] Mooney R. J., Roy L. Content-based book recommending using learning for text categorization. In proceedings of DL-00, 5[th] ACM Conference on Digital Libraries, pages 195-204, San Antonio, US, 2000. ACM Press, New York, US.
[22] Pazzani M., Billsus D. Learning and Revising User Profiles:The Identification of interesting Web Sites. Machine Learning, Vol. 27, 1997, p. 313-331.
[23] Pazzani M. A Framework for Collaborative, Content-Based, and Demographic Filtering. Artificial Intelligence Rev., pp. 393-408, Dec.1999.
[24] Pohl K., Böckle G., van der Linden F. Software Product Line Engineering – Foundations, Principles, and Techniques. Springer, Berlin, Heidelberg, New York, 2005.

[25] Riebisch M., Böllert K., Streitferdt D., Philippow I. Extending Feature Diagrams with UML Multiplicities. In Proceedings of the Sixth Conference on Integrated Design and Process Technology (IDPT 2002), Pasadena, CA, Juin 2002.

[26] Salinesi C., Mazo R., Diaz D., and Djebbi O. Solving Integer Constraint in Reuse Based Requirements Engineering. International Conference on Requirement Engineering (RE), Sydney, Australia, September 2010.

[27] Salton G. Automatic Text Processing: The Transformation, Analysis, and Retrieval of Information by Computer. Addison-Wesley, Reading, MA, 1989.

[28] Sarwar B., Karypis G., Konstan J., and Riedl J. Item-based collaborative filtering recommendation algorithms. In Proceedings of the 10th International WWW Conference, 2001.

[29] http://www.sei.cmu.edu/. SEI web page

[30] http://www.splot-research.org/

[31] Streitferdt D., Riebisch M., and Philippow I. Details of formalized relations in feature models using ocl. In ECBS, pages 297–304. IEEE Computer Society, 2003.

[32] Subbarayan S., Jensen R. M., Hadzic T., Andersen H. R., Hulgaard H., Moller J. Comparing two implementations of a complete and backtrack-free interactive con_gurator. In Workshop on CSP Techniques with Immediate Application (CSPIA), pages 97-111, 2004.

[33] Subbarayan S. Integrating CSP decomposition techniques and BDDs for compiling configuration problems. In International Conference on Integration of AI and OR Techniques in Constraint Programming for Combinatorial Optimization Problems (CPAIOR), volume 3524 of LNCS, pages 351-365. Springer, 2005.

[34] Van Deursen A., Klint P. Domain-Specific Language Design Requires Feature Descriptions. Journal of Computing and Information Technology, 10(1):1-17, 2002.

[35] Van Meteren R., Van Someren M. Using content-based filtering for recommendation. ECML 2000 workshop, 2000.

[36] White J., Dougherty B., Schmidt D.C. Selecting highly optimal architectural feature sets with filtered cartesian flattening. Journal of Systems and Software, 82(8):1268–1284, 2009.

[37] Astesana J-M., Cosserat L, Fargier H. Constraint-based vehicle configuration: a case study. In International Conference on Tools with Artificial Intelligence (ICTAI), IEEE Computer Society. Arras, 27-29 October 2010.